\documentclass[useAMS,usenatbib]{mn2e}
\usepackage[dvips]{graphicx}
\usepackage{amssymb}

\newcommand \nodata {...}
\newcommand \vcirc {V_\mathrm{circ}}
\newcommand \kms {\mathrm{km~s^{-1}}}

\title[Are fossil groups a challenge of the CDM paradigm?]{Are fossil
  groups a challenge of the Cold Dark Matter paradigm?}

\author[Zibetti, Pierini \& Pratt]{Stefano
  Zibetti$^{1}$\thanks{E-mail:
    zibetti@mpia.de}, Daniele Pierini$^{2}$, Gabriel W. Pratt$^{2}$ \\
  $^{1}$Max-Planck-Institut f\"ur Astronomie, K\"onigstuhl 17, D-69117 Heidelberg, Germany\\
  $^{2}$Max-Planck-Institut f\"ur extraterrestrische Physik, Postfach 1312, D-85741 Garching bei M\"unchen, Germany\\
}
\begin{document}
\bibliographystyle{mn2e}

\date{Accepted 2008 October 16. Received 2008 October 15; in original
  form 2008 July 15}

\pagerange{\pageref{firstpage}--\pageref{lastpage}} \pubyear{2008}

\maketitle

  \label{firstpage}

\begin{abstract}
  We study six groups and clusters of galaxies suggested in the
  literature to be `fossil' systems (i.e. to have luminous diffuse
  X-ray emission and a magnitude gap of at least 2 mag-R between the
  first and the second ranked member within half of the virial
  radius), each having good quality X-ray data and SDSS spectroscopic
  or photometric coverage out to the virial radius. The poor cluster
  AWM\,4 is clearly established as a fossil system, and we confirm the
  fossil nature of four other systems (RX\,J1331.5+1108,
  RX\,J1340.6+4018, RX\,J1256.0+2556 and RX\,J1416.4+2315), while the
  cluster RX\,J1552.2$+$2013 is disqualified as fossil system.  For
  all systems we present the luminosity functions within 0.5 and 1
  virial radius that are consistent, within the uncertainties, with
  the universal luminosity function of clusters. For the five {\em
    bona fide} fossil systems, having a mass range
  $2\times10^{13}-3\times10^{14}~\mathrm{M_\odot}$, we compute
  accurate cumulative substructure distribution functions (CSDFs) and
  compare them with the CSDFs of observed and simulated
  groups/clusters available in the literature. We demonstrate that the
  CSDFs of fossil systems are consistent with those of normal observed
  clusters and do not lack any substructure with respect to simulated
  galaxy systems in the cosmological $\mathrm{\Lambda}$CDM
  framework. In particular, this holds for the archetype fossil group
  RX\,J1340.6$+$4018 as well, contrary to earlier claims.

\end{abstract}


\section{Introduction}\label{intro_sec}
Early numerical simulations suggested that the most compact galaxy
groups could merge to form a single elliptical galaxy (hence a `fossil
group') in a few billion years \citep{barnes_89}.  An elliptical
galaxy formed by the merger of such a group retains its X-ray emitting
halo of hot gas, which is unaffected by merging
\citep{ponman_bertram_93}.  Following this indication,
\cite{ponman_etal_94} discovered the archetype fossil group
RX\,J1340.6$+$4018.

\cite{vikhlinin+99} and \cite{jones+03}, on the basis of ROSAT
observations, have suggested that fossil groups constitute a
considerable population of objects. Their X-ray extent, bolometric
X-ray luminosity ($L_{\mathrm{X, bol}} > 10^{42}
h_{50}^{-2}~\mathrm{erg}~\mathrm{s}^{-1}$), dark matter dominated
total mass, and mass in the diffuse hot gas component are comparable
to those of bright groups and poor clusters of galaxies ($\sim
10^{13}$--$10^{14}~h_{70}^{-1}~\mathrm{M}_{\sun}$). The brightest
member of a fossil group has an optical luminosity comparable to that
of a cluster cD galaxy (i.e., $M_\mathrm{R} < -22.5 + 5
\mathrm{log}~h_{50}$) and dominates the galaxy luminosity function of
the system. The observational definition of a fossil system lies in
the detection of extended, very luminous X-ray emission from the hot
gas of the intracluster medium (ICM), and in the existence of an
R-band magnitude gap $\Delta m_{12} > 2$ between the brightest and
second brightest members within $0.5 R_\mathrm{vir}$
\citep{jones_ponman_forbes_00}.

As noted by \cite{vikhlinin+99}, fossil groups may represent the
ultimate examples of hydrostatic equilibrium in virialised systems,
since they must have been undisturbed for a very long time if they are
the result of galaxy merging within a group.  High-resolution
hydrodynamical cosmological simulations in the $\mathrm{\Lambda}$CDM
framework have shown that fossil groups have already assembled half of
their final DM mass at redshifts $z \ge 1$, and subsequently they
typically grow by minor mergers only, whereas non-fossil systems of
similar masses on average form later \citep{donghia+05}.  The early
assembly of fossil groups leaves sufficient time for objects with
luminosities close to the characteristic luminosity (i.e., $L \sim
L^{\star}$) to merge into the central galaxy by dynamical friction,
producing the magnitude gap which defines a fossil system.  In
addition, the simulated fossil groups were found to be over-luminous
in X-rays relative to non-fossil groups of the same optical
luminosity, in qualitative agreement with observations
\citep[cf.][]{vikhlinin+99,jones_ponman_forbes_00}.  In a recent
paper, \cite{vonbendabeckmann+08} showed that many galaxy groups may
undergo a fossil phase in their lives but may not necessarily stay
fossil down to $z=0$, owing to renewed infall of $L \sim L^{\star}$
galaxies from the large-scale environment. Such infall episodes are
statistically more likely for more massive systems, so that the
fraction of quasi-fossil systems (i.e. those with a large luminosity
gap between the central galaxy and the most luminous satellite) is
lower among clusters than among groups
\citep{milosavljevic+06,yang+08}.

Fossil groups have become a puzzling problem to cosmology since
\citet[DL04]{donghia_lake_04} showed that, with respect to
state-of-the-art predictions on the frequency of substructures in cold
dark matter (CDM) halos \citep{delucia+04a}, a virialised system like
RX\,J1340.6$+$4018 lacks galaxies nearly as luminous as the Milky Way.
Conversely, the same numerical simulations are able to accurately
describe the frequency of substructures in galaxy clusters as massive
as Virgo or Coma to well below the circular velocity of a Milky
Way-size dark halo (i.e., with $\vcirc \le
220~\mathrm{km}~\mathrm{s}^{-1}$).  In this respect, fossil groups
appeared to exacerbate the so-called `small-scale crisis' of CDM
universes \citep{klypin+99b}. In fact, DL04 concluded that the missing
substructure problem affects systems up to the scale of groups
(typically with $\mathrm{k} T_\mathrm{X} \le 1~\mathrm{keV}$).
However, this result has been challenged by \cite{sales+07}, who find
that the abundance and luminosity function of simulated fossil systems
are in reasonable agreement with the few available observational
constraints.

Given the great interest of this cosmological issue, some optical
studies have aimed at better characterizing the mass and luminosity
function (LF) of already known fossil systems
\citep[e.g.][]{mendesdeoliveira+06,cypriano+06} or identifying new
ones \citep[e.g.][]{santos+07}, although identifying low-mass fossil
systems is hampered by the fact that groups are under-represented in
existing X-ray catalogs. In spite of the increasing quality of the
data and number of candidates, no new cumulative substructure
distribution function (CSDF)\footnote{The cumulative substructure
  distribution function gives the number of sub-halos with circular
  velocity $\vcirc$ larger than a given fraction of the circular
  velocity of the parent halo $V_\mathrm{parent}$.}  has so far been
produced for a fossil system to compare with that obtained by DL04 for
the archetype fossil group RX\,J1340.6$+$4018.  The determination of
the CSDF for fossil systems with a range of masses is the objective of
the present study.

In the following, we adopt a $\mathrm{\Lambda}$CDM cosmological model
($\Omega_{\mathrm{m}} = 0.3$, $\Omega_{\mathrm{\Lambda}} = 0.7$) with
$H_0 = 70~h_{70}^{-1}~\mathrm{km}~\mathrm{s}^{-1}~\mathrm{Mpc}^{-1}$.
This model is consistent with the main {\em WMAP5} results
\citep[cf.][]{hinshaw+08}.  This work is mainly based on Sloan Digital
Sky Survey \citep[SDSS]{SDSS} data from the sixth data release
\citep[DR6, and references therein]{DR6}.

%%%%%%%%%%%%%%%%%%%%%%%%%%%%%%%%%%%%%%%%%%%%%%%%%%%%%%%%%%%%%%%%%%%%%%%%%%%%%

\section{The fossil sample}\label{sample_sec}

In this study we consider five groups and poor clusters of galaxies
from the sample of fossil systems originally identified by
\cite{ponman_etal_94} and \cite{jones+03}.  In addition we include the
poor cluster AWM\,4, whose status as fossil system has been suggested
by \cite{lin_mohr_04} and is established in this work (see Appendix
\ref{sample_AWM4_sec}).  These six systems were selected to have (i)
high-quality X-ray observations, either with {\em Chandra}
\citep{weisskopf+00} or XMM-{\em Newton} \citep{jansen+01}, to allow
for a reliable determination of their halo properties (i.e.,
$R_{200}$, $M_{200}$, $V_\mathrm{parent}$)\footnote{$R_{200}$ is the
  radius within which the average density of the system is 200 times
  the critical density of the Universe. $M_{200}$ is the mass within
  $R_{200}$.}, and (ii) to be covered by SDSS photometry and, in the
case of AWM\,4, by SDSS spectroscopy.  This enables us to study
membership (via spectroscopic redshifts or combining photometric
redshifts and statistical foreground/background subtraction) and
photometric properties of members across a wide field, out to the
virial radius of a system.  Such coverage is mandatory to enable us to
compare the observed CSDFs with previous determinations and the output
of cosmological simulations in the literature, which are normally
computed out to the virial radius of a system (i.e.,
$R_{200}$)\footnote{Throughout the paper we use $R_{200}$ as a proxy
  for the virial radius $R_\mathrm{vir}$.}.  Table \ref{sample_tab}
lists coordinates and redshifts of the six systems.

\subsection{X-ray data}

The X-ray and derived halo properties are reported in Table
\ref{Xray_tab}, along with the references to the works from which
these data are taken.  For all systems (except AWM\,4 and
RX\,J1416.4$+$2315) $M_{200}$ and $R_{200}$ are computed from the
X-ray temperatures published by \cite{khosroshahi+07} using the
scaling relations given by \cite{arnaud+05}. We adopt their fits to
the $M_{200}-T_\mathrm{X}$ and $R_{200}-T_\mathrm{X}$ relations of the
entire sample to derive our fiducial estimates. In our analysis in
Sect. \ref{SDF_sec} we consider the effect of estimating mass and
radius from fits to both the entire sample of \cite{arnaud+05}, which
has a slope of 1.71, and from fits to the hot systems alone (${\mathrm
  k}T > 3.5$ keV), which has a slope of 1.49. The resulting masses are
reported in Table \ref{Xray_tab}. Our motivation for this choice of
relation is based on the recent results of \cite{sun+08}, who derive
the mass temperature scaling relation from 1 to 10 keV using high
quality Chandra data of a large sample of groups and clusters. Their
$M-T$ relation normalisation is similar to that of \cite{arnaud+05},
and they find a slope of 1.67, such that group masses would be 18 per
cent higher at 1 keV than that derived from the fit to the full sample
of \cite{arnaud+05}. In using both Arnaud et al relations we wish to
attempt to bracket the most likely value of the $M-T$ normalisation in
this mass range. We note however that \cite{khosroshahi+07} suggest
that fossil groups may fall low on the $M-T$ relation, although
\cite{sun+08} find no evidence for this with a smaller sample of
fossil groups. The extent of the offset found by Khosroshahi et
al. appears to be mass dependent and is negligible at higher group
temperatures (${\mathrm k}T \gtrsim 2$ keV), although it could be up
to a factor of 2-3 for the very lowest temperature systems. Deeper
X-ray observations of fossil groups have been obtained to help resolve
this issue, and will be the subject of
a forthcoming paper.\\
As we show in Sect. \ref{SDF_sec}, the CSDFs we derive are relatively
insensitive to the exact choice of X-ray scaling relation and our
conclusions are robust in case fossil groups should indeed fall low on
the $M-T$ relation.

For AWM\,4 and RX\,J1416.4$+$2315, estimates of $M_{200}$ and
$R_{200}$ are taken from \cite{gastaldello+07} and
\cite{khosroshahi+06}, respectively, who fitted NFW profiles
\citep{NFW} to the observed mass density profiles of these two
systems.  It is worth noting that there is a very good agreement
between the estimates given in the two papers referred above and those
derived from the scaling relations of \cite{arnaud+05}.  From Table
\ref{Xray_tab} one can clearly see that our sample spans a broad range
in mass and X-ray temperature, in the regime of groups and poor
clusters.

Once $M_{200}$ and $R_{200}$ are obtained, the circular velocity of
the parent halo of each system is computed :
\begin{equation}
V_\mathrm{parent}=\sqrt{\frac{G M_{200}}{R_{200}}}\label{vp_eq}
\end{equation}

\subsection{Optical data}\label{optdata_subs}

With the exception of AWM\,4, the nearest cluster at $z=0.0317$, the
analysis of all systems relies on SDSS photometric data only
(including photometric redshifts).  Although for some of them
spectroscopic redshifts are available through the SDSS or other
published catalogs, the lack of completeness or insufficient depth
prevents us from using them.  The SDSS spectroscopic data, in
particular, are limited to 17.77 mag \citep[$r$ band,
see][]{strauss_etal02}; this translates into absolute magnitude limits
ranging from $-20$ (for RX\,J1331.5$+$1108 at $z=0.081$) to -22.5 (for
RX\,J1256.0$+$2556 at $z=0.23$), thus making the faint end of the
luminosity function inaccessible to our analysis.  In contrast,
assuming a conservative limit of 20.5 $r$-mag, the SDSS photometry
allows us to probe down to roughly one magnitude fainter than
$L^{\star}$ for all systems. Although one could in principle combine
the complete information derived from photometric data alone with the
spectroscopic memberships for the brightest galaxies or in the regions
covered by other surveys, combining different selection functions is
an unnecessary complication. In particular, we have checked that for
all bright galaxies ($r<17.77$~mag) with available spectroscopy the
spectroscopic membership coincides with the purely photometric one.

In Appendix \ref{notesobj_sec} we comment further on issues specific
to individual systems and on their status of fossil. In summary,
AWM\,4 is established as fossil; RX\,J1331.5$+$1108, RX\,J1340.6+4018
and RX\,J1416.4+2315 are confirmed fossil systems; observational
uncertainties do not allow us to establish RX\,J1256.0+2556 as a
fossil with very high confidence, but we will consider it as a fossil
in the following; lastly, RX\,J1552.2$+$2013 does not match the
magnitude gap requirement and is disqualified as a fossil system.

\begin{table}
\begin{minipage}{\textwidth}
\caption{The sample}\label{sample_tab}
\begin{tabular}{lccc}
  \hline
  Denomination & RA & Dec  & z \\
                 & (J2000.0)  & (J2000.0) &        \\
  \hline

  AWM\,4           & 16:04:57.0 & +23:55:14 & 0.0317 \\
  RX\,J1256.0+2556 & 12:56:03.4 & +25:56:48 & 0.2320 \\
  RX\,J1331.5+1108 & 13:31:30.2 & +11:08:04 & 0.0810 \\
  RX\,J1340.6+4018 & 13:40:33.4 & +40:17:48 & 0.1710 \\
  RX\,J1416.4+2315 & 14:16:26.9 & +23:15:32 & 0.1370 \\
  RX\,J1552.2+2013 & 15:52:12.5 & +20:13:32 & 0.1350
\end{tabular}
\end{minipage}
\end{table}

\begin{table*}
\begin{minipage}{\textwidth}
\caption{X-ray and halo properties of the sample systems}\label{Xray_tab}
{\scriptsize
\begin{tabular}{lcrrrrrrrrrc}
  \hline
  Denomination & $k T_\mathrm{X}$  & $L_\mathrm{X}$\footnote{Bolometric, within $R_{200}$, unless specified otherwise.} & \multicolumn{2}{c}{$M_{200}$} & \multicolumn{2}{c}{$R_{200}$} & \multicolumn{2}{c}{$R_{200}$} & \multicolumn{2}{c}{$V_\mathrm{parent}$} & Ref.\footnote{G07=\cite{gastaldello+07}, K07=\cite{khosroshahi+07}, K06=\cite{khosroshahi+06}} \\
       & [$\mathrm{keV}$] & [$10^{42}~\mathrm{erg}~\mathrm{s}^{-1}$] & \multicolumn{2}{c}{[$10^{13}~\mathrm{M}_\odot$]} & \multicolumn{2}{c}{[$\mathrm{Mpc}$]}  & \multicolumn{2}{c}{[$\mathrm{arcmin}$]} & \multicolumn{2}{c}{[$\mathrm{km}~\mathrm{s}^{-1}$]} & \\

       &         &                    &                                 & `hot' &                      & `hot'  &                      & `hot' &                        & `hot'  &  \\

  \hline
  AWM\,4           & $2.48 \pm 0.06$ & 39.3\footnote{Within 455 kpc, (0.1--100 keV)} & $13.75\pm1.46$ & \nodata & $1.054\pm0.038$ & \nodata & $29.10\pm 1.05$ & \nodata & $783\pm 10$ & \nodata & G07\\
  RX\,J1256.0+2556 & $2.63 \pm 1.13$ & 50.0 & $15.75^{+13.38}_{-9.75}$ & 19.63 &  $1.034^{+0.234}_{-0.283}$ & 1.107 & $4.66^{+1.07}_{-1.28}$ & 4.99 &   $809\pm200$ & 873 & K07 \\
  RX\,J1331.5+1108 & $0.81 \pm 0.04$ &  2.1 &  $2.25\pm0.19$         & 3.67  & $0.571\pm0.016$         & 0.664 & $6.23\pm 0.17$       & 7.25 &  $411\pm 12$ & 487 & K07 \\

  RX\,J1340.6+4018 & $1.16 \pm 0.08$ &  5.2 &  $3.98^{+0.48}_{-0.46}$  & 5.99  & $0.670\pm 0.026$        & 0.759 & $3.83\pm 0.14$       & 4.35 &  $506\pm 20$ & 582 & K07\\
  RX\,J1416.4+2315 & $4.00 \pm 0.62$ &170.0 & $31.00\pm 10$          & \nodata & $1.220\pm 0.060$        & \nodata & $8.39\pm 0.41$       & \nodata & $1045\pm 90$ & \nodata & K06\\
  RX\,J1552.2+2013 & $2.85 \pm 0.90$ & 60.0 & $19.03^{+11.48}_{-9.12}$ & 23.28 & $1.139^{+0.193}_{-0.222}$ & 1.213 & $7.93^{+1.34}_{-1.55}$  & 8.45 & $849\pm 150$ & 909 & K07
\end{tabular}
}
\end{minipage}
\end{table*}

%%%%%%%%%%%%%%%%%%%%%%%%%%%%%%%%%%%%%%%%%%%%%%%%%%%%%%%%%%%%%%%%%%%%%%%%%%%%%

\section{The cumulative substructure distribution function: method}\label{SDF_sec}

In this section we describe how the cumulative substructure
distribution function is computed for the systems in our sample.  As
mentioned in Sect. \ref{intro_sec}, the CSDF of a galaxy system, $N(>
x\equiv\vcirc/V_\mathrm{parent})$, is defined as the number of members
(i.e., substructures of the parent DM halo), other than the brightest
member, which are satellites and have circular velocities $\vcirc$
larger than $x V_\mathrm{parent}$.  Here $V_\mathrm{parent}$ is the
circular velocity of the parent system considered as a whole (defined
in Eq. \ref{vp_eq}) and $x < 1$.  Theoretically, each galaxy in a
group/cluster is associated with a DM sub-halo of a given mass
$M_\mathrm{sub}$ and radius $R_\mathrm{sub}$ such that the circular
velocity is simply defined as $\vcirc=\sqrt{G
  M_\mathrm{sub}/R_\mathrm{sub}}$.  Observationally this information
is only accessible for few well studied galaxies. For the galaxies in
our systems circular velocities must be derived from basic observables
(luminosity, central velocity dispersion) by means of scaling
relations, as we describe in Sect. \ref{vcirc_sec}.

The second essential ingredient for computing the CSDF is the method
to count galaxies up to a given circular velocity, either by means of
a complete cluster membership selection based on spectroscopic
redshifts, or by means of a robust statistical subtraction of the
galaxy foreground/background from the counts in the cluster/group
region.  This is discussed in Sect. \ref{members_sec}.

\subsection{Inferring $\vcirc$}\label{vcirc_sec}

The structure and luminosity of galaxies are known to be linked to
their kinematics via scaling relations. For late-type (spiral)
galaxies the Tully-Fisher (TF) relation \citep{TF} links luminosity
and (maximum or asymptotic) circular velocity in the disk, which we
assume to coincide with $\vcirc$\footnote{This is an approximation
  that may lead to a systematic overestimate of the true $\vcirc$ at
  the sub-halo virial radius by some 10 per cent
  \citep[see][]{salucci+07}.}.  For early-type (elliptical) galaxies
the central velocity dispersion $\sigma_0$ can be inferred either from
the luminosity using the Faber-Jackson (FJ) relation \citep{FJ} or,
when more information about the surface brightness distribution is
available, via the fundamental plane (FP) relation that connects half
light radius, effective surface brightness, and central velocity
dispersion $\sigma_0$.  Once $\sigma_0$ is known, the circular
velocity can then be inferred.  In doing so, one can either assume
isothermal dynamics, in which case $\vcirc \approx \sqrt{2} \sigma_0$,
or adopt empirical scaling relations between the two quantities based
on very deep observations of extended rotation curves in early-type
galaxies \citep[e.g.][]{pizzella+05,ferrarese_02}.  In the following
analysis we adopt the relation given by \cite{pizzella+05}, viz.,
\begin{equation}
  \vcirc=(1.32\pm0.09)\sigma_0 + (46\pm 14) ~~~~[\mathrm{km~s^{-1}}]\label{pizzella_eq}
\end{equation}
as it includes a comprehensive sample of both low and high surface
brightness galaxies over the range between $\sim 50$ and 350 km
s$^{-1}$, and is therefore best suited for an heterogeneous sample
like ours.  In fact Equation \ref{pizzella_eq} applies to every
galaxy, irrespective of its morphological type, provided that the
central velocity dispersion is measured.

We note that the analysis presented in this work relies on the
reasonable assumption that scaling relations hold everywhere and
phenomena like the tidal truncation of sub-halos do not modify them
significantly.

In practice, we are faced with very different kinds of data, requiring
a diversity of approaches, as we now describe.

\subsubsection{$\vcirc$ in AWM\,4}\label{vcirc_AWM4_par}

Most of the galaxies in AWM\,4 have a measured velocity dispersion
from SDSS spectra.  For these galaxies we first correct the measured
value for the standard aperture corresponding to the effective radius
$r_e/8$, using the recipe given by \cite{joergensen+95}.  Then we
apply the scaling relation of \cite{pizzella+05} to infer $\vcirc$.

In cases where the velocity dispersion is not available, the
relatively small distance to this cluster allows a reliable estimate
of the structural parameters of the galaxies to be achieved.  Hence we
proceed as follows.

\begin{itemize}
\item[(i)] First we select early-type galaxies based on the
  concentration index $C_i \equiv
  R_{90~\mathrm{Petro},i}/R_{50~\mathrm{Petro},i} >2.5$ as given by
  the SDSS in $i$-band \citep[as in][]{bernardi+03_I}.  For these
  galaxies $\sigma_0$ is derived from the $r$-band FP relation of
  \cite{bernardi+03_III}, after applying $k$ and cosmological dimming
  $[(z+1)^4]$ corrections to the effective surface brightness.
  Thereafter $\sigma_0$ is converted into $\vcirc$ using the
  \cite{pizzella+05} relation as detailed above.

\item[(ii)] Bright galaxies (rest-frame $M_i<-19$ mag) with $C_i \le
  2.5$ are considered {\em bona fide} disc-dominated galaxies and
  their $\vcirc$ is computed via the TF relation. In order to minimise
  systematic errors, our best estimate of $\vcirc$ is given by the
  mean of the TF estimate of \citet[][using $i$-band Petrosian
  magnitudes]{pizagno+07} and of \citet[][using $i$-band `total'
  composite model magnitudes converted to $I_C$]{tully+98}. In both
  cases magnitudes are $k$-corrected and corrected for the
  inclination-dependent internal extinction using the recipe of
  \cite{tully+98}. $i$ and $I_C$ bands are explicitly chosen to
  minimise the amount and uncertainty of this correction
  \citep[cf. e.g.][]{pierini99,pierini+03}.

\item[(iii)] Finally, for galaxies fainter than $M_i=-19$ mag and with
  $C_i \le 2.5$, the concentration index is not indicative of the
  dynamical state (e.g. dwarf elliptical galaxies can be ``hot''
  systems yet have low concentration). In this case our best guess
  $\vcirc$ is given by the mean of the estimate from the FP -- as in
  case (i) -- and from the TF -- as in case (ii).
\end{itemize}

Statistical errors are computed by summing uncertainties on the
measured quantities propagated to $\vcirc$ and the typical
r.m.s. scatter around the scaling relations, in quadrature.  In
particular, we adopt a scatter of 15 $\kms$ for $\vcirc$ in the
\cite{pizzella+05} relation, 10\% r.m.s. for $\sigma_0$ from the FP,
and 15\% for $\vcirc$ about the TF relation \citep{pizagno+07}.  The
effect of random errors on the CSDF are quantified by means of Monte
Carlo simulations (see Sect. \ref{SDFsubsec}).

Systematic errors may have even greater impact on our conclusions
about the CSDFs as they can shift or expand/shrink the real
distributions. We consider systematic uncertainties for all adopted
scaling relations. In the FP and \cite{pizzella+05} relation we use
the uncertainties on the fitting parameters given in the reference
papers.  In particular, for the \cite{pizzella+05} relation we account
for the covariance between the two fitted coefficients by re-writing
the fitting function referred to the mean abscissa of the points and
considering only the error on the slope.  The covariance terms between
the FP parameters can then be safely neglected as Fig. 2 of
\cite{zibetti+02} illustrates.  For the TF estimates we adopt half of
the range spanned by the two relations of \cite{pizagno+07} and
\cite{tully+98} as a systematic uncertainty.

The set of all possible systematics and their permutations gives us
useful upper and lower limits for the computed CSDFs, as we will show
in Fig. \ref{SDFall}.

\subsubsection{$\vcirc$ for RX groups/clusters}\label{vcirc_RX_par}

For the five RX systems we rely on photometric data alone to compute
$\vcirc$ for the reasons outlined in Sect. \ref{sample_sec}.  In
addition, their higher redshift relative to AWM\,4 makes the use of
structural parameters much more uncertain.  Moreover, uncertainties on
membership determined through statistical methods (see
Sect. \ref{members_sec}) will completely dominate over the
uncertainties brought in by the scaling relations.  Therefore we
simplify the procedure described above.

For the RX systems we determine $\vcirc$ (i) either using the TF
relation of \cite{pizagno+07} for the most accurate $r$ band or (ii)
by converting $\sigma_0$, inferred from the $r$-band magnitude via the
FJ relation of \cite{bernardi+03_II}, into $\vcirc$ using the
\cite{pizzella+05} relation.  The choice between the two is dictated
by the colour of the galaxy, according to the well known colour
bimodality \citep[e.g.][]{baldry+04}.  For each cluster/group we look
at the $g-i$ colour distribution of galaxies with a photometric
redshift close to the redshift of the system, and determine {\em by
  eye} the cut between {\em blue} and {\em red} galaxies.  The values
of $\vcirc$ of the former are determined via the TF relation, while
for the latter we use the combination of FJ and \cite{pizzella+05}
relations.

All magnitudes we use for the scaling relations are taken from
SDSS-DR6 and are corrected for foreground Galactic extinction,
redshift \citep[$k$-corrections as given by][]{SDSS_photozs} and, for
red galaxies only, passive stellar evolution, according to the output
of the code of \cite{BC03}.  We have checked that our results are
largely insensitive to the particular choice of colour cut or,
alternatively, to a cut in spectral type, as derived from the
photometric redshift algorithm \citep[see][for details]{SDSS_photozs}.

\subsection{Membership and statistical foreground/background subtraction
  methods}\label{members_sec}

With the exception of AWM\,4, for which cluster membership is assigned
to individual galaxies via spectroscopic redshifts (see
Sect. \ref{sample_AWM4_sec}), member galaxy counts are based on
photometric redshifts from the SDSS-DR6 \citep{SDSS_photozs} and are
obtained after application of statistical foreground/background
subtraction methods.

For each system, we define a circle corresponding to the projected
$R_{200}$, centered at the coordinates of the peak of the X-ray
emission, and 1\,000 control fields of the same circular aperture,
randomly distributed inside an annulus with inner radius $2 R_{200}$
and outer radius 4 degrees.  The inner radius is chosen to avoid the
cluster region, while the outer radius allows us to avoid control
field overlapping.  We assume each galaxy lies at the redshift of the
system in question, and the corresponding property $X$ (e.g. absolute
magnitude, velocity etc.)  is computed.  Galaxy number distributions
for the property $X$ are computed in the cluster region ($D_0(X)$) and
in the control fields ($D_i(X)|_{\{i=1,1000\}}$).  The ``true''
distribution for the cluster is then given by $D_{\rmn
  {cluster}}(X)=D_0(X)-\overline{D(X)}$, where $\overline{D(X)}$ is
the median value of $D_i(X)$ over $i$ for any given value of
$X$.\footnote{The distribution computed in this way actually measures
  the {\em excess} of counts in the cluster vs. the field.  To obtain
  the true cluster counts, we correct the field counts by subtracting
  the contribution of the physical space that would be occupied by the
  cluster. Such a correction is well below the per cent level in all
  cases.}  The confidence range for $D_{\rmn{cluster}}(X)$ is given by
the distributions obtained by replacing the median value of $D_i(X)$
with the 16th and 84th percentile values.

Although in principle this method does not require any pre-selection
of galaxies, the background contamination is exceedingly high with
respect to the net signal contributed by the cluster, so that a
careful pre-selection is mandatory.  To this end we use photometric
redshift estimates and shape classifications provided by the SDSS for
every source.  Considering the distances to our systems and the
limiting magnitude of 20.5 $r$-mag (Petrosian, corrected for
foreground Galactic extinction), all cluster/group galaxies should be
safely resolved even in the worst case of a PSF FWHM equal to
1.5\arcsec.  Therefore we initially select only sources classified as
extended (``GALAXIES'').

We then apply a cut in photometric redshift to exclude the largest
possible number of contaminants in the fore- and background, but
avoiding exclusion of system members.  Unfortunately photometric
redshifts are typically affected by significant uncertainties, from
0.03 r.m.s. at 16 $r$-mag to 0.08 at 20.5 $r$-mag
\citep{SDSS_photozs}.  Therefore the selected ranges $z_{\rmn{sys}}\pm
\Delta z$ need to be quite broad.  We optimise the result by adapting
the cut widths $\Delta z$ to linearly scale as a function of apparent
magnitude, such that $\Delta z(m_r=14.5)=0.05$ and $\Delta
z(m_r=20.5)=0.1$.

This pre-selection in photometric redshift and stellarity index allows
us to attain a satisfactory signal-to-noise (S/N), while leaving us
with a clean and unbiased selection. This is confirmed by the
stability of our distributions when we adopt less restrictive cuts (at
the cost of a degradation in S/N however). Moreover, as already
mentioned in Sec. \ref{optdata_subs}, the memberships of the brightest
galaxies based on this photometric method coincide with those
resulting from spectroscopic redshifts, whenever these are
available. As a key test for our method, we next discuss the
luminosity functions for all of our systems.

\begin{figure*}

\includegraphics[width=\textwidth]{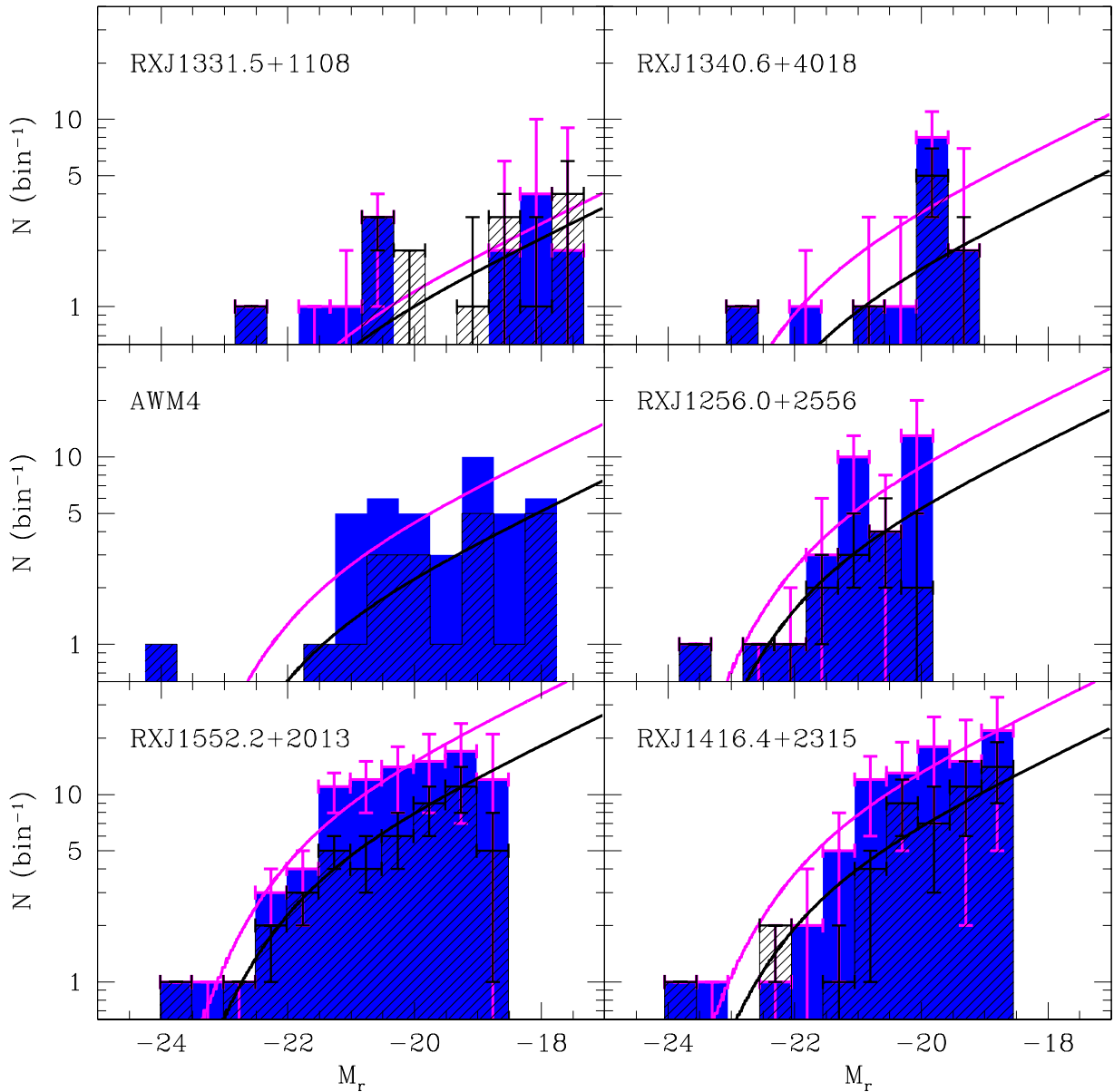}
\caption{The luminosity function of the six galaxy systems, ordered by
  ascending circular velocity from left to right and top to
  bottom. Counts are per bin of magnitude (0.5 mag width).  Filled
  blue histograms and magenta symbols represent the entire area within
  $R_{200}$, hatched histograms and black symbols represent the area
  within $0.5 R_{200}$. Error bars are derived via statistical methods
  (see text for details). The curves represent the analytical LF of
  Popesso et al. (2005a), normalised to the total number of galaxies
  down to 20 $r$-mag (apparent), with no other adjustable
  parameter.}\label{LFall}
\end{figure*}

%%%%%%%%%%%%%%%%%%%%%%%%%%%%%%%%%%%%%%%%%%%%%%%%%%%%%%%%%%%%%%%%%%%%%%%%%%%%%

\section{Results}

\subsection{The luminosity function of fossil systems}

Figure \ref{LFall} shows the luminosity function (LF) of the six
galaxy systems.  The objects are plotted from left to right and from
top to bottom in order of ascending circular velocity
$V_\mathrm{parent}$.  We examine the LF in two apertures, namely
within $R_{200}$ (blue filled histograms and magenta lines) and $0.5
R_{200}$ (black hatched histograms and black lines).  For all systems
but AWM\,4 error bars show the confidence intervals derived from
control field statistics.

%shape
The solid curves represent the universal LF of \cite{popesso+05_LF},
normalised to the total number of galaxies down to 20 $r$-mag
(apparent).  We stress that no other parameter is tuned in order to
improve the match between these curves and the measured histogram.
Errors are too large to claim any agreement or disagreement with the
universal LF for the least massive groups, because of the small number
statistics and the weak contrast against the background.  For the
three most massive systems, however, the plots show an almost perfect
agreement within the error bars\footnote{Note that the brightest
  galaxy is not considered as part of the regular luminosity
  function.}.

%total number of galaxies
\begin{figure}
  \includegraphics[width=0.5\textwidth]{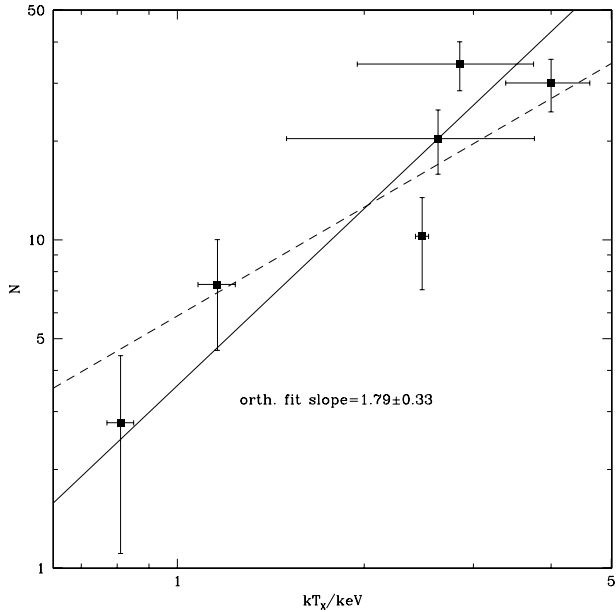}
  \caption{The ``galaxy richness'' of the systems $N$ versus the X-ray
    temperature. The solid line is the orthogonal least squares fit to
    the points, assuming poisson errors on the number counts. The
    slope is reported in the panels. The dashed line shows the slope
    of 1.1 for the analogous relation observed by Popesso et
    al. (2004).}\label{scaling_fig}
\end{figure}

The normalization $N$ of the LF\footnote{$N$ is the number of galaxies
  per 0.5 mag at $M_r=-18.0$.} scales, as expected, with cluster mass
or, equivalently, with X-ray temperature, as we show in
Fig. \ref{scaling_fig}. The orthogonal fit to the data points (solid
line) gives $N\propto L_{\mathrm X}^{1.79\pm0.33}$, where the error on
the exponent is formally derived under the hypothesis of gaussian
errors. If the assumption is made that the total optical luminosity of
the cluster ($r$-band), $L_\mathrm{opt,r}$, scales proportionally to
$N$, we can compare this result with the finding of
\cite{popesso+04_Xscale}: $L_\mathrm{opt,r}\propto L_{\mathrm
  X}^{1.1}$ (dashed line in Fig. \ref{scaling_fig}). Formally our
relation is significantly (at more than 2 sigmas) steeper than the one
found by \cite{popesso+04_Xscale}, but consistent with the 3/2 slope
expected for self-similar systems. However, our small number
statistics and possible systematics hidden behind the assumption
$L_\mathrm{opt,r}\propto N$ prevent us from drawing any quantitative
conclusion.

%gaps
The LFs within $0.5 R_{200}$ also confirm the magnitude gaps typical
of fossil systems (see Appendix \ref{notesobj_sec}). This is
particularly remarkable as it shows that the photometric redshift
pre-selection works very well in the bright regime, and provides very
high completeness and low contamination.

%cf spec LF from literature
Finally we note that the LF (inside $0.5 R_{200}$) for
RX\,J1416.4$+$2315 and RX\,J1552.2$+$2013 are in very good {\em
  quantitative} agreement with those determined by \cite{cypriano+06}
and \cite{mendesdeoliveira+06}, respectively, based on spectroscopic
membership determination.  A detailed comparison indicates that the
drop of number counts that we observe in RX\,J1552.2$+$2013 at
$M_r>-19$~mag is fully consistent with the observations of
\cite{mendesdeoliveira+06}, and therefore is most likely real.

In light of the above we conclude that our statistical method of
membership estimation is sufficiently robust and accurate to study the
LF, and hence is applicable also to study of the CSDF.

\subsection{Substructure Distribution Functions}\label{SDFsubsec}

\begin{figure*}
\includegraphics[width=\textwidth]{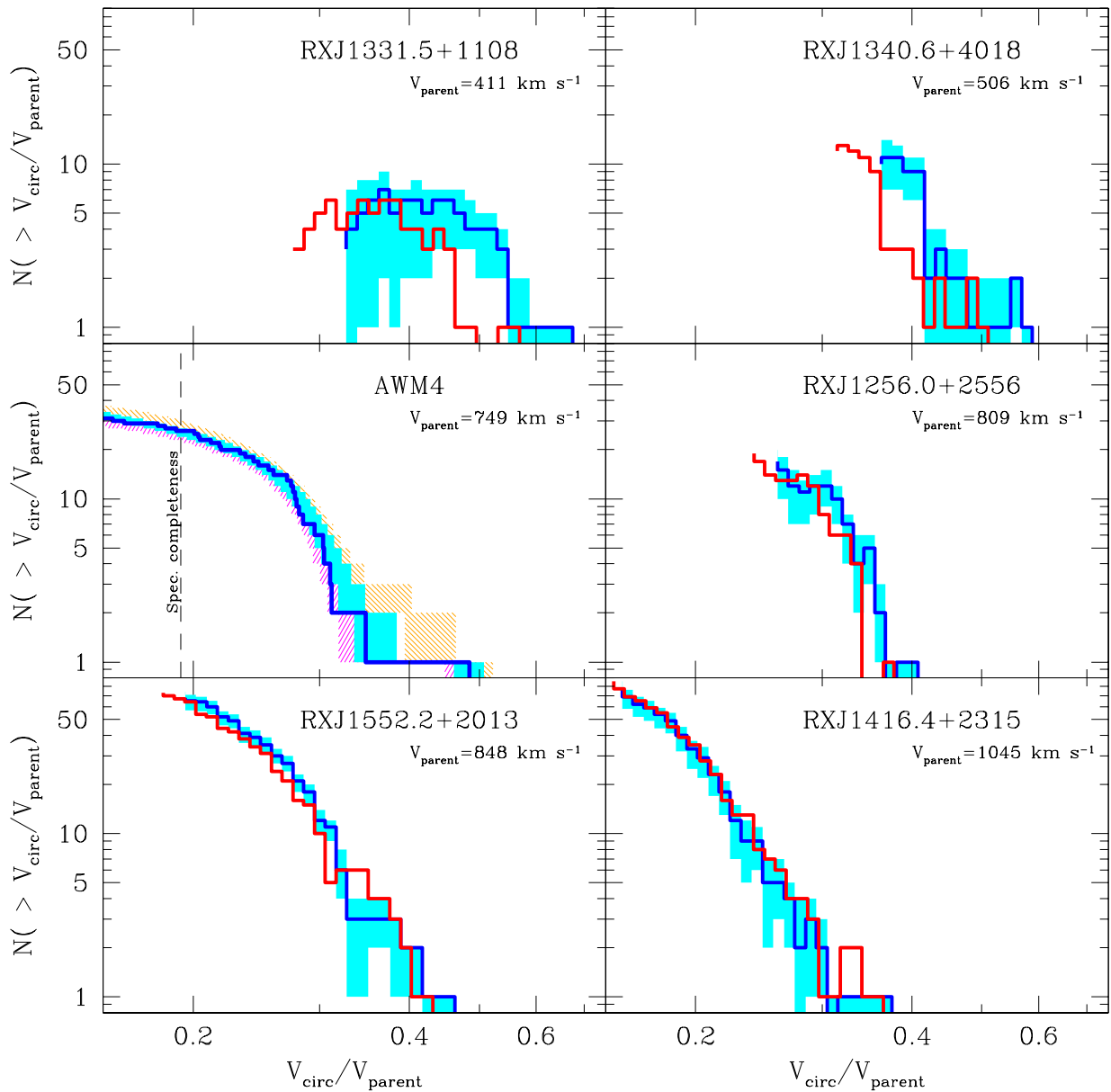}
\caption{The cumulative substructure distribution functions of the six
  galaxy systems, ordered by ascending circular velocity from left to
  right and top to bottom. The blue lines are our fiducial CSDF
  determinations. Cyan areas display statistical uncertainties. Red
  lines and hatched areas (for AWM\,4) show the effect of systematic
  uncertainties, as described in the text.}\label{SDFall}
\end{figure*}

Cumulative substructure distribution functions are constructed
following different methods for AWM\,4 and the five RX systems.  In
the case of AWM\,4 all spectroscopic members are used directly to
build the CSDF from their estimated $\vcirc/V_\mathrm{parent}$.  In
the case of RX systems, the method described in
Sect. \ref{members_sec} is adopted with the property $X$ replaced by
$\vcirc/V_\mathrm{parent}$.  The CSDFs obtained in this way, assuming
the values of $V_\mathrm{parent}$ given in Table \ref{Xray_tab}, are
shown as thick blue lines in Fig. \ref{SDFall}.

The limiting magnitudes that define the completeness of our sample
translate into a very complicated completeness function for
$\vcirc/V_\mathrm{parent}$ because of the different scaling relations
adopted.  By inverting each scaling relation we calculate the set of
$\vcirc$ corresponding to the limiting magnitudes of our samples.  The
maximum among these values is then our completeness limit for the
CSDF.  The CSDFs of the RX systems are cut at this limit in
Fig. \ref{SDFall}.  For AWM\,4 the entire CSDF is shown, and the
completeness limit is marked with a vertical dashed line.

Figure \ref{SDFall} also shows statistical and systematic
uncertainties.  Concerning AWM\,4, the $\vcirc$ of each galaxy has an
associated error.  To quantify the effect of these uncertainties on
the CSDF we run 10\,000 Monte Carlo simulations offsetting each
$\vcirc$ by a random $\Delta v$ drawn from a Gaussian distribution
with the same width as the associated uncertainty.  For each
realization the CSDF is computed.  The confidence range is determined
from the 16th--84th percentile range of $\vcirc/V_\mathrm{parent}$ at
each step of the CSDF, and is shown as the cyan shaded area in
Fig. \ref{SDFall}.  This randomisation procedure is undertaken
assuming the standard scaling relations.  If we now let the parameters
of the scaling relations vary within their uncertainties, we obtain
the orange and the magenta hatched regions as extreme cases.  On the
other hand, the X-ray data appear to be good enough to determine
$R_{200}$ and $V_\mathrm{parent}$ with an accuracy of better than 10
per cent. This implies that galaxy counts would change only slightly,
and the CSDF could shift horizontally by 10 per cent at most, which is
comparable to or less than the effect of other systematic
uncertainties.

For the five RX systems, we consider only uncertainties due to
statistical foreground/background subtraction and to the determination
of the properties of the parent system.  The cyan shaded areas in
Fig. \ref{SDFall} show the 16th--84th percentile range of variation
due to the statistical uncertainties of background counts.  The red
thick lines represent the CSDFs that we obtain by adopting the X-ray
scaling relations determined from systems with $k T > 3.5$ keV instead
of the full range (i.e., the parent system parameters derived from the
`hot' scaling relations, see Sect. \ref{sample_sec}).  The range of
variation due to each uncertainty is of the same magnitude at the
lower end of the mass range; at higher masses, uncertainties due to
statistical foreground/background subtraction dominate.

We note that the observed $M-T_\mathrm{X}$ relations exhibit
normalizations which are lower than those produced from simulations by
a factor of $\sim 10$ per cent, perhaps due to a neglect of
non-thermal pressure support \cite[]{arnaud+05,vikhlinin+06,nagai+07}.
Such a systematic effect would lead us to underestimate both $R_{200}$
and $V_\mathrm{parent}$ in the calculation of the CSDFs.  This would
tend to shift all CSDFs slightly down and to the right in
Fig. \ref{SDFall}. On the other hand, if fossil groups had lower
masses for a given $T_{\mathrm X}$ as claimed by
\cite{khosroshahi+07}, their CSDFs would shift to the right, in the
sense of more abundant substructure. This should mainly affect groups
with $\mathrm k T_{\mathrm X} \lesssim 2$ keV, i.e. only
RX\,J1331.5+1108 and RX\,J1340.5+4017 in our sample. It is worth
noting, however, that these two groups are those that lie closer to
the $M-T$ relation of normal groups \citep[see][their Figure
8]{khosroshahi+07}. By comparing $R_\delta$ and $M_\delta$ at
overdensity $\delta=500$ as given by \cite{khosroshahi+07} and as
estimated from our fiducial scaling relations \citep{arnaud+05}, we
conclude that the CSDFs of RX\,J1331.5+1108 and RX\,J1340.5+4017 could
further shift to higher $\vcirc/V_{\mathrm{parent}}$ by 25 and 17 per
cent respectively. Errors in the X-ray temperature itself propagate
quite weakly on the CSDF.  The effect of under(over)-estimating
$T_\mathrm{X}$ is to under(over)-estimate both $R_{200}$ and
$V_\mathrm{parent}$, with the result of decreasing (increasing) the
galaxy counts but simultaneously shifting the distribution to the
right (left).

%%%%%%%%%%%%%%%%%%%%%%%%%%%%%%%%%%%%%%%%%%%%%%%%%%%%%%%%%%%%%%%%%%%%%%%%%%%%%

\section{Discussion}

Figure \ref{SDFall} shows an interesting feature: the CSDFs shift to
lower $\vcirc/V_{\mathrm{parent}}$ by almost a factor of two as the
mass of the system increases. Noticeably this is in agreement with the
statistical behaviour of $V_{\mathrm{circ,2nd}}/V_{\mathrm{parent}}$
for the second brightest member of SDSS groups/clusters as a function
of mass as presented by \cite{yang+08}. From their Figures 6 and 7 we
compute that $V_{\mathrm{circ,2nd}}/V_{\mathrm{parent}}$ decreases by
$\approx 0.24$~dex when $M_{200}$ increases from $2\times10^{13}$ to
$3\times10^{14}~\mathrm{M_\odot}$, as in our sample. This decrease
nicely matches the shift observed in Fig. \ref{SDFall} and does not
point to any anomalous behaviour of fossil with respect to non-fossil
systems of the same mass.

As a test, following \cite{sales+07} we compute the magnitude gap
between the first and the tenth ranked member $\Delta M_{10}$ of
AWM\,4, using the spectroscopic membership determinations. With a
central galaxy luminosity $L=2.34\times 10^{11}\mathrm{L_\odot}$ and
$\Delta M_{10}=3.29$~mag, AWM\,4 falls in the region covered by the
simulations of isolated bright galaxies by \citet[their Figure
3]{sales+07}. Once more, this suggests that fossil systems are not
anomalous when compared to current $\mathrm{\Lambda}$CDM simulations.

\begin{figure*}
\includegraphics[angle=-90,width=\textwidth]{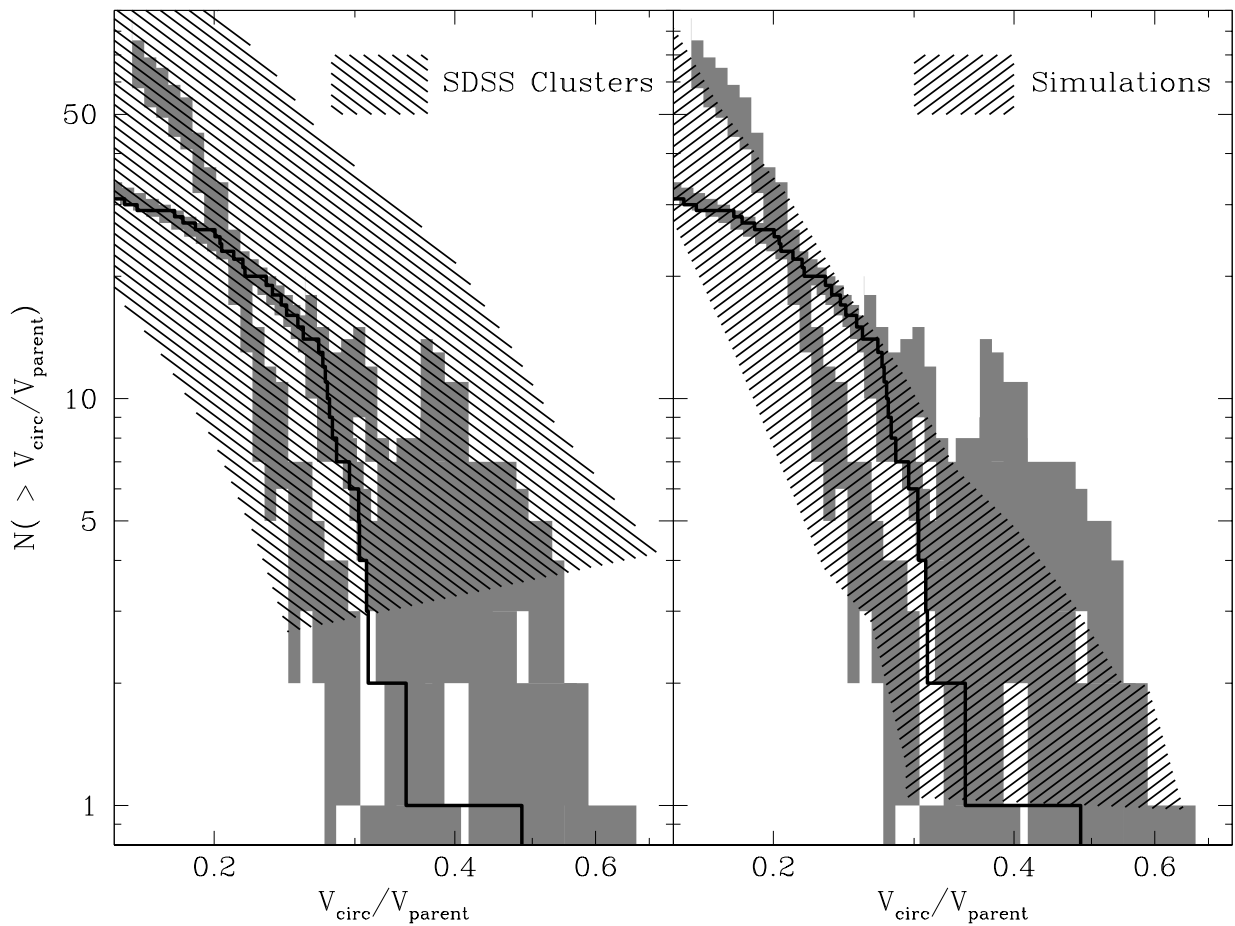}
\caption{The CSDFs of our five fossil systems (gray shaded areas,
  displaying the range of statistical uncertainties, same as cyan
  areas in Fig. 3) compared to CSDFs measured for 34 SDSS clusters of
  galaxies (hatched area, {\em left} panel) and simulated clusters
  (hatched area, {\em right} panel) from Desai et al. (2004). The CSDF
  of AWM\,4 is highlighted by the thick black line.}\label{SDFcomp}
\end{figure*}

To better prove this, we compare the CDFs of the five {\em bona fide}
fossil systems with those of observed nearby clusters/groups and of
simulated systems in the $\mathrm{\Lambda}$CDM framework.  In
particular we refer to the results presented in \citet[their Figure
5]{desai+04c}. These authors have measured CSDFs for 34
clusters/groups identified in the SDSS, with velocity dispersions
ranging from 250 to 1\,000 km s$^{-1}$. \cite{desai+04c} also consider
15 simulated groups/clusters in approximately the same velocity
dispersion range and measured the corresponding CSDFs. Figure
\ref{SDFcomp} reproduces their observed ({\em left} panel) and
simulated CSDFs ({\em right} panel) as hatched areas, overlaid onto
the CSDFs of our five fossil systems (grey-shaded areas, displaying
the range of statistical uncertainties, equivalent to the cyan area in
Fig. \ref{SDFall}; the thick black lines are the CSDF of AWM\,4).

The CSDFs of all fossil systems fall in the range of measured CSDFs
for SDSS clusters. In comparison to simulations the CSDFs of our
fossil systems are generally in good agreement, although an excess of
substructure is observed at high $\vcirc/V_\mathrm{parent}$. This
discrepancy is seen also between the real and the simulated clusters
by \cite{desai+04c}. As these authors show (their Section 6.4), it
most likely results from not properly taking the effects of baryons
into account in simulations. What is most interesting to note here,
however, is that none of the fossil systems appears to be lacking
significant amount of substructure, this result being robust against
systematic effects induced by different choices of scaling relations.

In particular, we are unable to reproduce the results of DL04
concerning RX\,1340.6$+$4018: although with large uncertainties, the
archetype fossil group does not show any evidence for missing
substructure. We note that the entire CSDF of DL04 (their Fig. 1) is
shifted to a factor of two lower velocities with respect to ours.
This might result from a different choice of scaling relations,
although (as we show for AWM\,4) choices within a reasonable range of
parameters cannot change the results by such a large amount.  Moreover
our photometric dataset allows us to apply scaling laws in a very
accurate and galaxy-type specific way, as detailed in
Sect. \ref{vcirc_RX_par}.  In contrast to DL04 we also take advantage
of a much better and more complete sample selection, based on complete
photometric coverage over the entire virialised region of the group,
and perform robust statistical foreground/background subtraction.
Finally, we have thoroughly investigated all possible sources of
uncertainty which could propagate into our determination of the CSDFs
of these systems, finding them all to be relatively small.

In conclusion, the present analysis rules out lack of substructure in
fossil clusters ($M_{200}\gtrsim 10^{14}~\mathrm{M_\odot}$), where our
two photometrically derived CSDFs are robust and consistent with the
CSDF of AWM\,4 based on spectroscopy. The same appears to hold for
lower mass fossil systems, although with non-negligible uncertainties.
Spectroscopic determinations of CSDFs for fossil groups
($M_{200}\approx10^{13}-10^{14}~\mathrm{M_\odot}$) would greatly help
to reach a firm conclusion in this mass range.

%%%%%%%%%%%%%%%%%%%%%%%%%%%%%%%%%%%%%%%%%%%%%%%%%%%%%%%%%%%%%%%%%%%%%%%%%%%%%

\section{Conclusions}

We have studied six galaxy groups and clusters suggested in the
literature to be fossil systems, and having high quality X-ray and
SDSS spectroscopic or photometric information. Among them, we have
established AWM\,4 as fossil system. We confirm three other systems
(RX\,J1331.5$+$1108, RX\,J1340.6+4018 and RX\,J1416.4+2315) to be
fossil. RX\,J1256.0+2556 has characteristics very close to a genuine
fossil system, although observational errors are still too large for a
robust classification as fossil. Finally we demonstrate that
RX\,J1552.2$+$2013 does not match the magnitude gap requirement to
qualify as a fossil.

For each system we have computed the luminosity functions within 0.5
and 1 virial radius. Although with uncertainties, that are remarkably
large for the least massive systems, they are consistent with the
universal luminosity function of clusters derived by
\cite{popesso+04_Xscale}.

We have derived detailed cumulative substructure distribution
functions (CSDFs) for the six galaxy groups and clusters. Our
motivation was to produce CSDFs for systems with a range of masses in
the group and poor cluster regime, for comparison with the original
CDSF derived for the archetype fossil system RX\,J1340.6$+$4018 by
\citet[DL04]{donghia_lake_04}. In fact, these authors claimed a lack
of substructure for RX\,J1340.6$+$4018, suggesting that the so-called
`small-scale crisis' of Milky-Way-size haloes \citep{klypin+99b}
exists up to the scale of X-ray bright groups. In addition we wanted
to compare the CSDFs of our fossil systems with results from normal
groups and clusters identified in the SDSS \citep{desai+04c},
and in cosmological simulations \citep{desai+04c,sales+07}.\\
Our conclusions are as follows: \vspace{-0.2cm}
\begin{itemize}
\item The CSDF of AWM\,4, based on spectroscopic data, is completely
  consistent both with \cite{desai+04c}'s data and with simulations.
  AWM\,4 also matches the simulations by \cite{sales+07} of isolated
  bright galaxies in terms of central galaxy luminosity vs. magnitude
  gap between the first and the tenth ranked member.
\item The photometrically derived CSDFs of the other four {\em bona
    fide} fossil systems are all completely consistent with the CSDF
  envelope derived from observed normal groups and clusters
  \citep{desai+04c}. With respect to numerically simulated systems in
  \cite{desai+04c} none of our fossil systems appears to lack any
  substructure.
\end{itemize}
\vspace{-0.2cm}

We therefore conclude that no evidence can be provided that the
`small-scale crisis' is occurring on the scale of fossil systems.  In
other words, the presence of a large magnitude gap between the first
and second ranked members of a group/cluster does not imply anything
special about its substructure, which can be fully accounted for by
existing LCDM simulations.

Interestingly, we also observe a systematic shift of the CSDFs toward
lower $\vcirc/V_\mathrm{parent}$ at increasing $V_\mathrm{parent}$.
This can be reproduced using the scaling relations of first and second
brightest members vs. halo mass derived for a representative sample of
groups and clusters by \cite{yang+08}. Once more, this reinforces the
idea that fossil systems are not anomalous as far as scaling relations
are concerned.

Further deep, wide-field spectroscopic observations of fossil systems
are required to confirm the results we have obtained in the present
work.  This is particularly necessary at the low end of the mass
range.  On the theoretical side, analyses which mimic observational
approaches such as that described here would allow us to compare real
data and simulations on an equal footing. This paper is the first in
a series which will characterise many of these aspects of fossil
groups.

\section*{Acknowledgements}
We thank Simon White and Hans-Walter Rix for useful comments.
S.Z. acknowledges the hospitality of the Max-Planck-Institut f\"ur
extraterrestrische Physik in Garching.  D.P. acknowledges useful
discussions with E. D'Onghia and the hospitality of the Institute for
Theoretical Physics of the University of Z\"urich.  D.P. acknowledges
support from the German \emph{Deut\-sches Zen\-trum f\"ur Luft- und
  Raum\-fahrt, DLR\/} project number 50~OR~0405.  G.W.P acknowledges
support from DfG Transregio Programme TR~33.

Funding for the SDSS and SDSS-II has been provided by the Alfred
P. Sloan Foundation, the Participating Institutions, the National
Science Foundation, the U.S. Department of Energy, the National
Aeronautics and Space Administration, the Japanese Monbukagakusho, the
Max Planck Society, and the Higher Education Funding Council for
England. The SDSS Web Site is http://www.sdss.org/.\\
The SDSS is managed by the Astrophysical Research Consortium for the
Participating Institutions. The Participating Institutions are the
American Museum of Natural History, Astrophysical Institute Potsdam,
University of Basel, University of Cambridge, Case Western Reserve
University, University of Chicago, Drexel University, Fermilab, the
Institute for Advanced Study, the Japan Participation Group, Johns
Hopkins University, the Joint Institute for Nuclear Astrophysics, the
Kavli Institute for Particle Astrophysics and Cosmology, the Korean
Scientist Group, the Chinese Academy of Sciences (LAMOST), Los Alamos
National Laboratory, the Max-Planck-Institute for Astronomy (MPIA),
the Max-Planck-Institute for Astrophysics (MPA), New Mexico State
University, Ohio State University, University of Pittsburgh,
University of Portsmouth, Princeton University, the United States
Naval Observatory, and the University of Washington.

This research has made use of the NASA/IPAC Extragalactic Database
(NED) which is operated by the Jet Propulsion Laboratory, California
Institute of Technology, under contract with the National Aeronautics
and Space Administration.

\bibliography{MN-08-1176-MJ_R1}

\begin{thebibliography}{}

\bibitem[\protect\citeauthoryear{{Adelman-McCarthy} \&
  {et~al.}}{{Adelman-McCarthy} et~al.}{2008}]{DR6}
{Adelman-McCarthy} J.~K., {et~al.}, 2008, \apjs, 175, 297

\bibitem[\protect\citeauthoryear{{Albert}, {White} \& {Morgan}}{{Albert}
  et~al.}{1977}]{1977ApJ...211..309A}
{Albert} C.~E., {White} R.~A., {Morgan} W.~W., 1977, \apj, 211, 309

\bibitem[\protect\citeauthoryear{{Arnaud}, {Pointecouteau} \& {Pratt}}{{Arnaud}
  et~al.}{2005}]{arnaud+05}
{Arnaud} M., {Pointecouteau} E., {Pratt} G.~W., 2005, \aap, 441, 893

\bibitem[\protect\citeauthoryear{{Baldry}, {Glazebrook}, {Brinkmann},
  {Ivezi{\'c}}, {Lupton}, {Nichol} \& {Szalay}}{{Baldry}
  et~al.}{2004}]{baldry+04}
{Baldry} I.~K.,  {Glazebrook} K.,  {Brinkmann} J.,  {Ivezi{\'c}} {\v Z}.,
  {Lupton} R.~H.,  {Nichol} R.~C.,    {Szalay} A.~S.,  2004, \apj, 600, 681

\bibitem[\protect\citeauthoryear{{Barnes}}{{Barnes}}{1989}]{barnes_89}
{Barnes} J.~E.,  1989, \nat, 338, 123

\bibitem[\protect\citeauthoryear{{Bernardi} \&
  {et~al.}}{{Bernardi} {et~al.}}{2003a}]{bernardi+03_I}
{Bernardi} M., {et~al.}, 2003a, \aj, 125, 1817

\bibitem[\protect\citeauthoryear{{Bernardi} \&
  {et~al.}}{{Bernardi} {et~al.}}{2003b}]{bernardi+03_II}
{Bernardi} M., {et~al.}, 2003b, \aj, 125, 1849

\bibitem[\protect\citeauthoryear{{Bernardi} \&
  {et~al.}}{{Bernardi} {et~al.}}{2003c}]{bernardi+03_III}
{Bernardi} M., {et~al.}, 2003c, \aj, 125, 1866

\bibitem[\protect\citeauthoryear{{Bernardi}, {Hyde}, {Sheth}, {Miller} \&
  {Nichol}}{{Bernardi} et~al.}{2007}]{bernardi+07}
{Bernardi} M., {Hyde} J.~B., {Sheth} R.~K., {Miller} C.~J., {Nichol} R.~C.,
  2007, \aj, 133, 1741

\bibitem[\protect\citeauthoryear{{Bruzual} \&
  {Charlot}}{{Bruzual} \& {Charlot}}{2003}]{BC03}
{Bruzual} G., {Charlot} S., 2003, \mnras, 344, 1000

\bibitem[\protect\citeauthoryear{{Csabai}, {Budav{\'a}ri}, {Connolly},
  {Szalay}, {Gy{\H o}ry}, {Ben{\'{\i}}tez}, {Annis}, {Brinkmann}, {Eisenstein},
  {Fukugita}, {Gunn}, {Kent}, {Lupton}, {Nichol} \& {Stoughton}}{{Csabai}
  et~al.}{2003}]{SDSS_photozs}
{Csabai} I.,  {Budav{\'a}ri} T.,  {Connolly} A.~J.,  {Szalay} A.~S.,  {Gy{\H
  o}ry} Z.,  {Ben{\'{\i}}tez} N.,  {Annis} J.,  {Brinkmann} J.,  {Eisenstein}
  D.,  {Fukugita} M.,  {Gunn} J.,  {Kent} S.,  {Lupton} R.,  {Nichol} R.~C.,
  {Stoughton} C.,  2003, \aj, 125, 580

\bibitem[\protect\citeauthoryear{{Cypriano}, {Mendes de Oliveira} \&
  {Sodr{\'e}}}{{Cypriano} et~al.}{2006}]{cypriano+06}
{Cypriano} E.~S.,  {Mendes de Oliveira} C.~L.,    {Sodr{\'e}} L.~J.,  2006,
  \aj, 132, 514

\bibitem[\protect\citeauthoryear{{De Lucia}, {Kauffmann}, {Springel}, {White},
  {Lanzoni}, {Stoehr}, {Tormen} \& {Yoshida}}{{De Lucia}
  et~al.}{2004}]{delucia+04a}
{De Lucia} G.,  {Kauffmann} G.,  {Springel} V.,  {White} S.~D.~M.,  {Lanzoni}
  B.,  {Stoehr} F.,  {Tormen} G.,    {Yoshida} N.,  2004, \mnras, 348, 333

\bibitem[\protect\citeauthoryear{{Desai}, {Dalcanton}, {Mayer}, {Reed},
  {Quinn}, \& {Governato}}{{Desai} et~al.}{2004}]{desai+04c}
{Desai} V., {Dalcanton} J.~J., {Mayer} L., {Reed} D., {Quinn} T.,
  {Governato} F., 2004, \mnras, 351, 265

\bibitem[\protect\citeauthoryear{{de Vaucouleurs}, {de Vaucouleurs}, {{Corwin},
  Jr.}, {Buta}, {Paturel} \& {Fouque}}{{de Vaucouleurs} et~al.}{1991}]{RC3}
{de Vaucouleurs} G.,  {de Vaucouleurs} A.,  {{Corwin}, Jr.} H.~G.,  {Buta}
  R.~J.,  {Paturel} G.,    {Fouque} P.,  1991, {Third Reference Catalogue of
  Bright Galaxies}.
Volume 1-3, XII, 2069 pp.~7 figs..~ Springer-Verlag Berlin Heidelberg New York

\bibitem[\protect\citeauthoryear{{D'Onghia} \& {Lake}}{{D'Onghia} \&
  {Lake}}{2004}]{donghia_lake_04}
{D'Onghia} E.,  {Lake} G.,  2004, \apj, 612, 628

\bibitem[\protect\citeauthoryear{{D'Onghia}, {Sommer-Larsen}, {Romeo},
  {Burkert}, {Pedersen}, {Portinari} \& {Rasmussen}}{{D'Onghia}
  et~al.}{2005}]{donghia+05}
{D'Onghia} E.,  {Sommer-Larsen} J.,  {Romeo} A.~D.,  {Burkert} A.,  {Pedersen}
  K.,  {Portinari} L.,    {Rasmussen} J.,  2005, \apjl, 630, L109

\bibitem[\protect\citeauthoryear{{Faber} \& {Jackson}}{{Faber} \&
  {Jackson}}{1976}]{FJ}
{Faber} S.~M.,  {Jackson} R.~E.,  1976, \apj, 204, 668

\bibitem[\protect\citeauthoryear{{Ferrarese}}{{Ferrarese}}{2002}]{ferrarese_02}
{Ferrarese} L.,  2002, \apj, 578, 90

%\bibitem[\protect\citeauthoryear{{Finoguenov}, {Reiprich} \&
%  {B{\"o}hringer}}{{Finoguenov} et~al.}{2001}]{finoguenov+01}
%{Finoguenov} A.,  {Reiprich} T.~H.,    {B{\"o}hringer} H.,  2001, \aap, 368,
%  749

\bibitem[\protect\citeauthoryear{{Gastaldello}, {Buote}, {Humphrey},
  {Zappacosta}, {Bullock}, {Brighenti} \& {Mathews}}{{Gastaldello}
  et~al.}{2007}]{gastaldello+07}
{Gastaldello} F.,  {Buote} D.~A.,  {Humphrey} P.~J.,  {Zappacosta} L.,
  {Bullock} J.~S.,  {Brighenti} F.,    {Mathews} W.~G.,  2007, \apj, 669, 158

\bibitem[\protect\citeauthoryear{{Gastaldello}, {Buote}, {Brighenti} \&
  {Mathews}}{{Gastaldello} et~al.}{2008}]{gastaldello+08}
{Gastaldello} F.,  {Buote} D.~A.,  {Brighenti} F.,    {Mathews} W.~G.,  2008,
  \apjl, 673, L17

\bibitem[\protect\citeauthoryear{{Gavazzi}, {Zibetti}, {Boselli}, {Franzetti},
  {Scodeggio} \& {Martocchi}}{{Gavazzi} et~al.}{2001}]{dEvirgo}
{Gavazzi} G.,  {Zibetti} S.,  {Boselli} A.,  {Franzetti} P.,  {Scodeggio} M.,
   {Martocchi} S.,  2001, \aap, 372, 29

\bibitem[\protect\citeauthoryear{{Hinshaw} \&
  {et~al.}}{{Hinshaw} {et~al.}}{2008}]{hinshaw+08}
{Hinshaw} G., {et~al.}, 2008, ArXiv e-prints, 803

\bibitem[\protect\citeauthoryear{{Jansen}, {Lumb}, {Altieri}, {Clavel}, {Ehle},
  {Erd}, {Gabriel}, {Guainazzi}, {Gondoin}, {Much}, {Munoz}, {Santos},
  {Schartel}, {Texier} \& {Vacanti}}{{Jansen} et~al.}{2001}]{jansen+01}
{Jansen} F.,  {Lumb} D.,  {Altieri} B.,  {Clavel} J.,  {Ehle} M.,  {Erd} C.,
  {Gabriel} C.,  {Guainazzi} M.,  {Gondoin} P.,  {Much} R.,  {Munoz} R.,
  {Santos} M.,  {Schartel} N.,  {Texier} D.,    {Vacanti} G.,  2001, \aap, 365,
  L1

\bibitem[\protect\citeauthoryear{{Jarrett}, {Chester}, {Cutri}, {Schneider},
  {Skrutskie} \& {Huchra}}{{Jarrett} et~al.}{2000}]{2MASS}
{Jarrett} T.~H.,  {Chester} T.,  {Cutri} R.,  {Schneider} S.,  {Skrutskie} M.,
    {Huchra} J.~P.,  2000, \aj, 119, 2498

\bibitem[\protect\citeauthoryear{{Jeltema}, {Mulchaey}, {Lubin} \&
  {Fassnacht}}{{Jeltema} et~al.}{2007}]{jeltema+07}
{Jeltema} T.~E.,  {Mulchaey} J.~S.,  {Lubin} L.~M.,    {Fassnacht} C.~D.,
  2007, \apj, 658, 865

\bibitem[\protect\citeauthoryear{{Jones} \& {Forman}}{{Jones} \&
  {Forman}}{1999}]{1999ApJ...511...65J}
{Jones} C.,  {Forman} W.,  1999, \apj, 511, 65

\bibitem[\protect\citeauthoryear{{Jones}, {Ponman} \& {Forbes}}{{Jones}
  et~al.}{2000}]{jones_ponman_forbes_00}
{Jones} L.~R.,  {Ponman} T.~J.,    {Forbes} D.~A.,  2000, \mnras, 312, 139

\bibitem[\protect\citeauthoryear{{Jones}, {Ponman}, {Horton}, {Babul},
  {Ebeling} \& {Burke}}{{Jones} et~al.}{2003}]{jones+03}
{Jones} L.~R.,  {Ponman} T.~J.,  {Horton} A.,  {Babul} A.,  {Ebeling} H.,
  {Burke} D.~J.,  2003, \mnras, 343, 627

\bibitem[\protect\citeauthoryear{{J{\o}rgensen}, {Franx} \&
  {Kjaergaard}}{{J{\o}rgensen} et~al.}{1995}]{joergensen+95}
{J{\o}rgensen} I.,  {Franx} M.,    {Kjaergaard} P.,  1995, \mnras, 276, 1341

\bibitem[\protect\citeauthoryear{{Khosroshahi}, {Maughan}, {Ponman} \&
  {Jones}}{{Khosroshahi} et~al.}{2006}]{khosroshahi+06}
{Khosroshahi} H.~G.,  {Maughan} B.~J.,  {Ponman} T.~J.,    {Jones} L.~R.,
  2006, \mnras, 369, 1211

\bibitem[\protect\citeauthoryear{{Khosroshahi}, {Ponman} \&
  {Jones}}{{Khosroshahi} et~al.}{2007}]{khosroshahi+07}
{Khosroshahi} H.~G.,  {Ponman} T.~J.,    {Jones} L.~R.,  2007, \mnras, 377, 595

\bibitem[\protect\citeauthoryear{{Klypin}, {Kravtsov}, {Valenzuela} \&
  {Prada}}{{Klypin} et~al.}{1999}]{klypin+99b}
{Klypin} A.,  {Kravtsov} A.~V.,  {Valenzuela} O.,    {Prada} F.,  1999, \apj,
  522, 82

\bibitem[\protect\citeauthoryear{{Koranyi} \& {Geller}}{{Koranyi} \&
  {Geller}}{2002}]{koranyi_geller_02}
{Koranyi} D.~M.,  {Geller} M.~J.,  2002, \aj, 123, 100

\bibitem[\protect\citeauthoryear{{Lin} \& {Mohr}}{{Lin} \&
  {Mohr}}{2004}]{lin_mohr_04}
{Lin} Y.-T.,  {Mohr} J.~J.,  2004, \apj, 617, 879

\bibitem[\protect\citeauthoryear{{Mendes de Oliveira}, {Cypriano} \&
  {Sodr{\'e}}}{{Mendes de Oliveira} et~al.}{2006}]{mendesdeoliveira+06}
{Mendes de Oliveira} C.~L.,  {Cypriano} E.~S.,    {Sodr{\'e}} L.~J.,  2006,
  \aj, 131, 158

\bibitem[\protect\citeauthoryear{{Milosavljevi{\'c}}, {Miller}, {Furlanetto} \&
  {Cooray}}{{Milosavljevi{\'c}} et~al.}{2006}]{milosavljevic+06}
{Milosavljevi{\'c}} M.,  {Miller} C.~J.,  {Furlanetto} S.~R.,    {Cooray} A.,
  2006, \apjl, 637, L9

\bibitem[\protect\citeauthoryear{{Morgan}, {Kayser} \& {White}}{{Morgan}
  et~al.}{1975}]{1975ApJ...199..545M}
{Morgan} W.~W.,  {Kayser} S.,    {White} R.~A.,  1975, \apj, 199, 545

\bibitem[\protect\citeauthoryear{{Nagai}, {Kravtsov}, \&
  {Vikhlinin}}{{Nagai} et al.}{2007}]{nagai+07}
{Nagai} D., {Kravtsov} A.~V., {Vikhlinin} A., 2008, \apj, 668, 1

\bibitem[\protect\citeauthoryear{{Navarro}, {Frenk} \& {White}}{{Navarro}
  et~al.}{1997}]{NFW}
{Navarro} J.~F.,  {Frenk} C.~S.,    {White} S.~D.~M.,  1997, \apj, 490, 493

\bibitem[\protect\citeauthoryear{{O'Sullivan}, {Vrtilek}, {Kempner}, {David} \&
  {Houck}}{{O'Sullivan} et~al.}{2005}]{2005MNRAS.357.1134O}
{O'Sullivan} E.,  {Vrtilek} J.~M.,  {Kempner} J.~C.,  {David} L.~P.,    {Houck}
  J.~C.,  2005, \mnras, 357, 1134

\bibitem[\protect\citeauthoryear{{Pfeffermann}, {Briel}, {Hippmann},
  {Kettenring}, {Metzner}, {Predehl}, {Reger}, {Stephan}, {Zombeck}, {Chappell}
  \& {Murray}}{{Pfeffermann} et~al.}{1987}]{pfeffermann+87}
{Pfeffermann} E.,  {Briel} U.~G.,  {Hippmann} H.,  {Kettenring} G.,  {Metzner}
  G.,  {Predehl} P.,  {Reger} G.,  {Stephan} K.-H.,  {Zombeck} M.,  {Chappell}
  J.,    {Murray} S.~S.,  1987, in {Koch} E.-E.,  {Schmahl} G.,  eds, Soft
  X-ray optics and technology; Proceedings of the Meeting, Berlin, Federal
  Republic of Germany, Dec. 8-11, 1986, Bellingham, WA, Society of
  Photo-Optical Instrumentation Engineers, Volume 733, 1987, p. 519. Vol.~733
  of Presented at the Society of Photo-Optical Instrumentation Engineers (SPIE)
  Conference, {The focal plane instrumentation of the ROSAT Telescope}.
pp 519--+

\bibitem[\protect\citeauthoryear{{Pierini}}{Pierini}{1999}]{pierini99}
{Pierini} D., 1999, \aap, 352, 49

\bibitem[\protect\citeauthoryear{{Pierini}, {Gordon}, \&
  {Witt}}{{Pierini} {et al.}}{2003}]{pierini+03}
{Pierini} D., {Gordon} K.~D., {Witt} A.~N., 2003, RMxAC, 17, 200

\bibitem[\protect\citeauthoryear{{Pizagno}, {Prada}, {Weinberg}, {Rix},
  {Pogge}, {Grebel}, {Harbeck}, {Blanton}, {Brinkmann} \& {Gunn}}{{Pizagno}
  et~al.}{2007}]{pizagno+07}
{Pizagno} J.,  {Prada} F.,  {Weinberg} D.~H.,  {Rix} H.-W.,  {Pogge} R.~W.,
  {Grebel} E.~K.,  {Harbeck} D.,  {Blanton} M.,  {Brinkmann} J.,    {Gunn}
  J.~E.,  2007, \aj, 134, 945

\bibitem[\protect\citeauthoryear{{Pizzella}, {Corsini}, {Dalla Bont{\`a}},
  {Sarzi}, {Coccato} \& {Bertola}}{{Pizzella} et~al.}{2005}]{pizzella+05}
{Pizzella} A.,  {Corsini} E.~M.,  {Dalla Bont{\`a}} E.,  {Sarzi} M.,  {Coccato}
  L.,    {Bertola} F.,  2005, \apj, 631, 785

\bibitem[\protect\citeauthoryear{{Ponman} \& {Bertram}}{{Ponman} \&
  {Bertram}}{1993}]{ponman_bertram_93}
{Ponman} T.~J.,  {Bertram} D.,  1993, \nat, 363, 51

\bibitem[\protect\citeauthoryear{{Ponman}, {Allan}, {Jones}, {Merrifield},
  {McHardy}, {Lehto} \& {Luppino}}{{Ponman} et~al.}{1994}]{ponman_etal_94}
{Ponman} T.~J.,  {Allan} D.~J.,  {Jones} L.~R.,  {Merrifield} M.,  {McHardy}
  I.~M.,  {Lehto} H.~J.,    {Luppino} G.~A.,  1994, \nat, 369, 462

\bibitem[\protect\citeauthoryear{{Popesso}, {B{\"o}hringer}, {Brinkmann},
  {Voges} \& {York}}{{Popesso} et~al.}{2004}]{popesso+04_Xscale}
{Popesso} P.,  {B{\"o}hringer} H.,  {Brinkmann} J.,  {Voges} W.,    {York}
  D.~G.,  2004, \aap, 423, 449

\bibitem[\protect\citeauthoryear{{Popesso}, {B{\"o}hringer}, {Romaniello} \&
  {Voges}}{{Popesso} et~al.}{2005}]{popesso+05_LF}
{Popesso} P.,  {B{\"o}hringer} H.,  {Romaniello} M.,    {Voges} W.,  2005,
  \aap, 433, 415

\bibitem[\protect\citeauthoryear{{Sales}, {Navarro}, {Lambas}, {White} \&
  {Croton}}{{Sales} et~al.}{2007}]{sales+07}
{Sales} L.~V.,  {Navarro} J.~F.,  {Lambas} D.~G.,  {White} S.~D.~M.,
  {Croton} D.~J.,  2007, \mnras, 382, 1901

\bibitem[\protect\citeauthoryear{{Salucci}, {Lapi}, {Tonini}, {Gentile},
  {Yegorova} \& {Klein}}{{Salucci} et~al.}{2007}]{salucci+07}
{Salucci} P.,  {Lapi} A.,  {Tonini} C.,  {Gentile} G.,  {Yegorova} I.,
  {Klein} U.,  2007, \mnras, 378, 41

\bibitem[\protect\citeauthoryear{{Santos}, {Mendes de Oliveira} \&
  {Sodr{\'e}}}{{Santos} et~al.}{2007}]{santos+07}
{Santos} W.~A.,  {Mendes de Oliveira} C.,    {Sodr{\'e}} L.~J.,  2007, \aj,
  134, 1551

\bibitem[\protect\citeauthoryear{{Strauss} \&
  {et~al.}}{{Strauss} {et~al.}}{2002}]{strauss_etal02}
{Strauss} M.~A., {et~al.}, 2002, \aj, 124, 1810

\bibitem[\protect\citeauthoryear{{Sun}, {Voit}, {Donahue}, {Jones} \&
  {Forman}}{{Sun} et~al.}{2008}]{sun+08}
{Sun} M., {Voit} G.~M., {Donahue} M., {Jones} C., {Forman} W., 2008,
  arXiv0805.2320

\bibitem[\protect\citeauthoryear{{Tully} \& {Fisher}}{{Tully} \&
  {Fisher}}{1977}]{TF}
{Tully} R.~B.,  {Fisher} J.~R.,  1977, \aap, 54, 661

\bibitem[\protect\citeauthoryear{{Tully}, {Pierce}, {Huang}, {Saunders},
  {Verheijen} \& {Witchalls}}{{Tully} et~al.}{1998}]{tully+98}
{Tully} R.~B.,  {Pierce} M.~J.,  {Huang} J.-S.,  {Saunders} W.,  {Verheijen}
  M.~A.~W.,    {Witchalls} P.~L.,  1998, \aj, 115, 2264

\bibitem[\protect\citeauthoryear{{Vikhlinin}, {McNamara}, {Hornstrup},
  {Quintana}, {Forman}, {Jones} \& {Way}}{{Vikhlinin}
  et~al.}{1999}]{vikhlinin+99}
{Vikhlinin} A.,  {McNamara} B.~R.,  {Hornstrup} A.,  {Quintana} H.,  {Forman}
  W.,  {Jones} C.,    {Way} M.,  1999, \apjl, 520, L1

\bibitem[\protect\citeauthoryear{{Vikhlinin}, {Kravtsov}, {Forman}, {Jones},
  {Markevitch}, {Murray}, \& {Van Speybroeck}}{{Vikhlinin}
  et~al.}{2006}]{vikhlinin+06}
{Vikhlinin} A., {Kravtsov} A., {Forman} W., {Jones} C., {Markevitch} M.,
  {Murray} S.~S., {Van Speybroeck} L., 2006, \apj, 640, 691

\bibitem[\protect\citeauthoryear{{Vikhlinin}, {Burenin}, {Ebeling}, {Forman},
  {Hornstrup}, {Jones}, {Kravtsov}, {Murray}, {Nagai}, {Quintana},
  {Voevodkin}}{{Vikhlinin} et~al.}{2008}]{vikhlinin+08}
{Vikhlinin} A., {Burenin} R.~A., {Ebeling} H., {Forman} W.~R.,
  {Hornstrup} A., {Jones} C., {Kravtsov} A.~V., {Murray} S.~S.,
  {Nagai} D., {Quintana} H., {Voevodkin} A. 2008, arXiv0805.2207

\bibitem[\protect\citeauthoryear{{von Benda-Beckmann}, {D'Onghia},
  {Gottl{\"o}ber}, {Hoeft}, {Khalatyan}, {Klypin} \& {M{\"u}ller}}{{von
  Benda-Beckmann} et~al.}{2008}]{vonbendabeckmann+08}
{von Benda-Beckmann} A.~M.,  {D'Onghia} E.,  {Gottl{\"o}ber} S.,  {Hoeft} M.,
  {Khalatyan} A.,  {Klypin} A.,    {M{\"u}ller} V.,  2008, \mnras, 386, 2345

\bibitem[\protect\citeauthoryear{{Wegner}, {Colless}, {Saglia}, {McMahan},
  {Davies}, {Burstein} \& {Baggley}}{{Wegner}
  et~al.}{1999}]{1999MNRAS.305..259W}
{Wegner} G.,  {Colless} M.,  {Saglia} R.~P.,  {McMahan} R.~K.,  {Davies} R.~L.,
   {Burstein} D.,    {Baggley} G.,  1999, \mnras, 305, 259

\bibitem[\protect\citeauthoryear{{Weisskopf}, {Tananbaum}, {Van Speybroeck} \&
  {O'Dell}}{{Weisskopf} et~al.}{2000}]{weisskopf+00}
{Weisskopf} M.~C.,  {Tananbaum} H.~D.,  {Van Speybroeck} L.~P.,    {O'Dell}
  S.~L.,  2000, in {Truemper} J.~E.,  {Aschenbach} B.,  eds, Proc. SPIE Vol.
  4012, p. 2-16, X-Ray Optics, Instruments, and Missions III, Joachim E.
  Truemper; Bernd Aschenbach; Eds. Vol.~4012 of Presented at the Society of
  Photo-Optical Instrumentation Engineers (SPIE) Conference, {Chandra X-ray
  Observatory (CXO): overview}.
pp 2--16

\bibitem[\protect\citeauthoryear{Yang, Mo, 
\& van den Bosch}{2008}]{yang+08} Yang X., Mo H.~J., van den Bosch F.~C., 2008, ApJ, 676, 248 

\bibitem[\protect\citeauthoryear{{York} \&
  {et~al.}}{{York} {et~al.}}{2000}]{SDSS}
{York} D.~G., {et~al.}, 2000, \aj, 120, 1579

\bibitem[\protect\citeauthoryear{{Zibetti}, {Gavazzi}, {Scodeggio}, {Franzetti}
  \& {Boselli}}{{Zibetti} {et~al.}}{2002}]{zibetti+02}
{Zibetti} S., {Gavazzi} G., {Scodeggio} M., {Franzetti} P., {Boselli} A.,
  2002, \apj, 579, 261

\end{thebibliography}
%\begin{thebibliography}{}
%
%\bibitem[\protect\citeauthoryear{Abazajian et 
%al.}{2003}]{DR1} Abazajian K., et al., 2003, AJ, 126, 2081
%
%\end{thebibliography}

\appendix

\section{Notes on individual objects}\label{notesobj_sec}

\subsection{AWM\,4}\label{sample_AWM4_sec}

AWM\,4 belongs to a special subset of poor clusters, originally
selected in the optical by \citet*[MKW]{1975ApJ...199..545M} and
\citet*[AWM]{1977ApJ...211..309A} in a search for cD-like galaxies
outside rich clusters, where these galaxies had traditionally been
found.  NGC\,6051, $V_\mathrm{T}^0= 12.93$~mag \citep{RC3}, located at
$z=0.031755 \pm 0.000033$ \citep{1999MNRAS.305..259W}, is the cD of
AWM\,4.  \cite{koranyi_geller_02} subsequently identified 28 members
brighter than $\mathrm{R}=15.5$~mag.  Most are early-type galaxies,
concentrated in the center and with a smooth Gaussian velocity
distribution centered at the velocity of NGC\,6051.  The velocity
dispersion of the system is 440 km s$^{-1}$.

AWM\,4 first appeared as an extended X-ray source in the catalogue of
images of galaxy clusters detected by the {\em Einstein} Imaging
Proportional Counter \citep{1999ApJ...511...65J}.  No substructure nor
a departure from symmetry was found in this X-ray image of AWM\,4 at a
24 arcsec-resolution.  This has been confirmed by the latest X-ray
observations, with XMM-{\em Newton} and at a resolution of 6 arcsec
\citep{2005MNRAS.357.1134O}, where the X-ray emission is centred on
NGC\,6051.  Despite the presence of a powerful AGN \citep[see][for a
thorough discussion]{gastaldello+08}, the relaxed appearance of AWM\,4
both in optical and X-rays motivated \cite{gastaldello+07} to include
this system in their sample of 16 bright relaxed groups/clusters to
which they applied a hydrostatic analysis to measure mass profiles.
We use the results of their best fitting NFW profile to characterise
the DM halo of AWM\,4.  The resulting total mass, $1.6\times10^{14}
\mathrm{M_\odot}$, and X-ray temperature, 2.5 keV, qualify AWM\,4 as a
poor cluster.

\cite{lin_mohr_04} first pointed out that AWM\,4 appears to be a
fossil system, since it is X-ray luminous and meets the magnitude-gap
criterion in Ks band\footnote{Photometry in the J, H, and Ks bands,
  complete down to $\mathrm{Ks} = 13.5$, is available from the Two
  Micron All-Sky Survey extended source catalogue
  \citep[2MASS,][]{2MASS}.}.  Within a half of the projected virial
radius from the X-ray center of AWM\, two spectroscopic members have
the same Ks-band magnitude gap from NGC\,6051, of $\Delta m=2.2$~mag.
However, fossil groups were defined using R-band photometry
\citep{jones+03}.  In the following we show that the gap is present in
$R$ band too.  AWM\,4 is covered by SDSS imaging and spectroscopy out
to its virial radius.  Only 4 over 135 spectroscopic targets
\citep[see][for the spectroscopic target selection in the
SDSS]{strauss_etal02} do not have a valid redshift measurement down to
the limit of $r=17.77$~mag (corrected for Galactic extinction),
corresponding to the absolute magnitude $\mathrm{M}_r = -17.91$~mag at
the redshift of AWM\,4 for the average $k$-correction.  The
spectroscopic completeness reaches 100\% at $r=17.25$~mag; it
decreases to 90\% between 17.25 and 17.77~mag.  This allows us to
perform a complete cluster-member identification that is almost a
factor 10 deeper in luminosity than previously done by
\cite{koranyi_geller_02}.  Moreover, the superior photometric quality
of the SDSS images allows us to establish the status of AWM\,4 as
fossil system, by safely assuming that the gap in $r$ band is the same
as in R band.

We identify as members all galaxies within the projected virial radius
with spectroscopic redshift that differs from the velocity of the
system by less than 1\,000 km s$^{-1}$.  We find a total of 42
members, 23 within half of the virial radius (including NGC\,6051).
As SDSS photometry is known to underestimate the luminosity of large
galaxies \citep[see, e.g.,][]{bernardi+07} because of the problematic
sky subtraction inherent in the automatic pipeline, we perform our own
photometric measurements on the original $r$-band SDSS ``corrected
frames'' for all galaxies within $0.5 R_{200}$, following the same
procedure described in \cite{dEvirgo}.  The sky background is measured
in empty areas, sufficiently far away from galaxies, after masking
stars.  Elliptical isophotes are fitted to the images of individual
galaxies using the IRAF task {\em ellipse}, after carefully masking
contaminating sources, in particular the bright star projected on top
of NGC\,6051.  The resulting 1-D azimuthally-averaged surface
brightness profiles are fitted with analytical functions (exponential,
de Vaucouleurs or combined).  Finally, total magnitudes are derived by
extrapolating the measured flux to infinity, according to the best
fitting analytic model.  As expected, the luminosities of the
brightest galaxies are significantly underestimated by the SDSS
automatic pipeline.  By comparing the so-called `model' magnitudes
from the SDSS-DR6 with our own measurements, we find that NGC\,6051 is
reported 0.67 magnitude too faint, while other galaxies down to
$\approx 15$~mag are systematically fainter by 0.1 to 0.2 magnitude.

Magnitudes for the three brightest members of AWM\,4 are reported in
Table \ref{photo_BCMs}.  All magnitudes are corrected for foreground
Galactic extinction.  Column 4 lists our extrapolated magnitudes,
reported along with the statistical uncertainty due to photometric
errors and fitting uncertainties, and upper and lower systematic
shifts due to background uncertainties.  Column 5 contains our
magnitude determinations within the isophotal ellipse corresponding to
25 mag arcsec$^{-2}$.  Column 6 and 7 are the Petrosian and model
magnitudes from the SDSS, respectively.

\begin{table*}
\begin{minipage}{\textwidth}
\caption{Photometry of the three brightest member galaxies in AWM\,4}\label{photo_BCMs}
\begin{tabular}{ccccccc}
  \hline
  Rank & RA & Dec  & mag$_\infty$ & mag$_{\mu25}$ & Petrosian & model \\
  & (J2000.0)  & (J2000.0) \\
  (1) & (2) & (3)  & (4) & (5) & (6) & (7) \\
  \hline
  1 (cD, NGC\,6051) & 16:04:56.79 & +23:55:56.4 & $11.92 \pm 0.11 ^{+0.21}_{-0.27}$ & 12.46 & 12.72 & 12.60\\
  2                 & 16:05:17.61 & +23:45:20.4 & $14.15 \pm 0.03 ^{+0.02}_{-0.02}$ & 14.27 & 14.44 & 14.37\\
  3                 & 16:04:50.55 & +23:58:29.6 & $14.73 \pm 0.03 ^{+0.02}_{-0.02}$ & 14.77 & 14.83 & 14.70
\end{tabular}
\end{minipage}
\end{table*}

From our extrapolated magnitudes we conclude that
AWM\,4 fulfills the requirement of magnitude gap $\ge 2$~mag in $r$ band
($\approx$ R band), with a $\Delta m_{12}=2.23\pm0.11^{+0.21}_{-0.27}$~mag.
Comparison with the other magnitude determinations
and the systematic uncertainty due to the background level also demonstrate
how critical the inclusion of the diffuse envelope of the cD is
in establishing AWM\,4 as a genuine fossil system.

\subsection{RX\,J1256.0$+$2556}\label{sample_RXJ1256_sec}

RX\,J1256.0$+$2556 is one of the fossil groups originally included in
the sample of \cite{jones+03}.  These authors reported that the
presence of the magnitude gap within $0.5 R_\mathrm{vir}$ critically
depends on the exact value of $R_\mathrm{vir}$.  Although a better
determination is now available from \cite{khosroshahi+07}, the
uncertainty on the X-ray temperature of this system ($2.63\pm
1.13$~keV) translates into a large uncertainty on
$R_{200}=1.03^{+0.24}_{-0.28}
\mathrm{Mpc}=4.66^{+1.08}_{-1.26}~\mathrm{arcmin}$.  The galaxy
[JML2007] J125557.90+255819.6 \citep[whose membership is confirmed
by][]{jeltema+07} is only 1.2 mag fainter than the brightest member
and is located at 1.96\arcmin~from the peak of the X-ray emission.
All other potential members with distance $\le 1.96$\arcmin~have a
magnitude difference with respect to the cD in excess of 2 mag.  The
uncertainty on $R_{200}$ therefore does not allow us to establish
whether the $\ge 2$ mag gap is present or not inside $0.5 R_{200}$.

Although \cite{jeltema+07} have conducted a spectroscopic campaign on
RX\,J1256.0$+$2556, we can not rely on their data to study the CSDF of
this system due to the relatively high incompleteness of their survey
(45--90\% at $\mathrm{V}=20.5$), and especially the limited span of
their data (the inner 700 kpc, thus only 70\% of $R_{200}$).  On the
other hand, SDSS-DR6 only provides sparse spectroscopic redshifts in
this region.  Therefore, for the following analysis we will rely on
SDSS photometric data alone.

\subsection{RX\,J1331.5$+$1108}\label{sample_RXJ1331_sec}

This group has the lowest temperature (0.81 keV) in our sample.  From
available SDSS spectroscopy and photometry we find $\Delta
m_{12}=1.93$~mag (SDSS model magnitudes), with the second brightest
member within $0.5 R_{200}$ (SDSS J133141.49+110644.6) being located
at 3.07\arcmin$=(0.493\pm0.013) R_{200}$ from the X-ray peak. As for
AWM\,4, we perform our own photometric analysis and find a gap $\Delta
m_{12}=2.17~$mag instead. The disagreement with SDSS is possibly
generated by sky oversubtraction for the brightest galaxy in the SDSS
photometric pipeline. Thus we confirm this group as a genuine fossil.

\subsection{RX\,J1340.6$+$4018}\label{sample_RXJ1340_sec}

This group is the fossil group archetype \citep{ponman_etal_94}.  The
magnitude gap is unambiguous here: the second brightest extended
object within $0.5 R_{200}$ is $\approx 2.6$ mag fainter than the cD.
DL04 based their claim for a substructure crisis in fossil groups
scale on data from this object.  Contrary to DL04, who approximated
the circular velocity of this group with
$V_\mathrm{parent}=\sqrt\sigma$ \citep[$\sigma$ being the 1D projected
velocity dispersion of the group reported
by][]{jones_ponman_forbes_00}, we use more accurate X-ray
determinations, but obtain a very similar value, within 30 km
s$^{-1}$.

\subsection{RX\,J1416.4$+$2515}\label{sample_RXJ1416_sec}

This system, with its $M_{200}\approx 3\times
10^{14}~\mathrm{M_\odot}$, is rather a poor cluster than a group.  It
has recently been studied by \cite{cypriano+06}, who obtained
spectroscopic redshift and studied the luminosity function in the
inner $\approx 3$\arcmin~($\approx 0.35 R_{200}$).  They confirm a
magnitude gap of $\Delta m_{12}=16.73-14.19=2.54$~mag in $i$ band
(AB), which we can safely assume to be close enough to the gap in $R$
band.  Although the region within $0.5 R_{200}$ is not completely
covered by their survey, we check against SDSS imaging that no
brighter member is missed.  It is worth noting, once again, that the
SDSS magnitude for the cD is severely under-estimated by 0.54 $i$-mag
and therefore would not have allowed a reliable estimation of the
magnitude gap to be obtained.

The limited coverage provided by the observations in \cite{cypriano+06}
forces us to rely on SDSS photometry to compute the CSDF.

\subsection{RX\,J1552.2$+$2013}\label{sample_RXJ1552_sec}

Similarly to RX\,J1416.4$+$2515, RX\,J1552.2$+$2013 is a poor cluster
and has been the target of a spectroscopic campaign aimed at
determining membership and LF \citep{mendesdeoliveira+06}.  As in the
previous case, the coverage is limited to the inner $\approx
3$\arcmin~($\approx 0.4 R_{200}$).  \cite{mendesdeoliveira+06} report
a gap $\Delta m_{12}\approx 2.2$~mag in $i$ band (AB; see their
Fig. 2).  However, by inspecting a larger region using SDSS images, at
2.6\arcmin~from the X-ray peak we find an elliptical galaxy
(SDSS\,J155201.61$+$201350.5) of 16.0 $i$-mag, that is only 1.2 mag
fainter than the brightest cluster galaxy.  This galaxy is missing
from the sample of \cite{mendesdeoliveira+06}, likely because is only
partly included in their image (their Fig. 1).  For this galaxy, SDSS
gives a photometric redshift of 0.15, hence fully consistent with the
redshift of the cluster.  Assuming that this redshift is correct and
our determination of $R_{200}$ is correct, even if only within the large
uncertainties quoted in Table \ref{Xray_tab}, this cluster does not
qualify as fossil.

\label{lastpage}
\end{document}